\documentclass[10pt,twocolumn,letterpaper]{article}

\usepackage{cvpr}
\usepackage{times}
\usepackage{epsfig}
\usepackage{transparent}
\usepackage{graphicx}
\usepackage[export]{adjustbox}
\usepackage{array}
\usepackage{amsmath}
\usepackage{amsthm}
\usepackage{amssymb}
\usepackage[noend]{algpseudocode}
\usepackage{algorithm,lipsum}
\usepackage{tikz}
\usepackage{multicol}
\usepackage{caption}
\usepackage{subcaption}
\usepackage{pgfplots}
\pgfplotsset{compat=newest}
\usepackage{tcolorbox}
\usepackage{dblfloatfix}

\usepackage{import}
\usetikzlibrary{quotes,arrows.meta}
\usetikzlibrary{positioning}

\def\edgecolor{rgb:blue,4;red,1;green,4;black,3}
\newcommand{\midarrow}{\tikz \draw[-Stealth,line width =0.8mm,draw=\edgecolor] (-0.3,0) -- ++(0.3,0);}

\usepackage{Ball}
\usepackage{Box}
\usepackage{RightBandedBox}

\usetikzlibrary{positioning}
\usetikzlibrary{3d} 

\def\ConvColor{rgb:yellow,5;red,2.5;white,5}

\def\PoolColor{rgb:red,1;black,0.3}
\def\UnpoolColor{rgb:blue,2;green,1;black,0.3}

\def\SoftmaxColor{rgb:magenta,5;black,7}

\usepackage[breaklinks=true,bookmarks=false]{hyperref}

\newtheorem{theorem}{Theorem}

\newcommand{\myindent}[1]{
\newline\makebox[#1cm]{}
}
\algnewcommand{\IIf}[1]{\State\algorithmicif\ #1\ \algorithmicthen}

\cvprfinalcopy 

\usepackage{color}
\usepackage[normalem]{ulem}
\usepackage{url}

\usepackage[belowskip=-5pt,aboveskip=10pt]{caption}
\DeclareUrlCommand\ULurl{
  
  }

\usepackage{authblk}

\author{{\large \normalfont Matan Rusanovsky$^{1,2}$, Gal Oren$^{1,3}$, Sigalit Ifergane$^{4}$ and Ofer Beeri$^{4}$}\newline
{\large \normalfont $^1$ Department of Computer Science, Ben-Gurion University of the Negev, Be'er Sheva, Israel}\newline
{\large \normalfont $^2$ Israel Atomic Energy Commission, Tel Aviv, Israel}\newline
{\large \normalfont $^3$ Department of Physics, Nuclear Research Center-Negev, Be'er-Sheva, Israel}\newline
{\large \normalfont $^4$ Department of Materials, Nuclear Research Center-Negev, Be'er-Sheva, Israel}\newline
{\tt\small \{matanru,orenw\}@post.bgu.ac.il},
{\tt\small \{ofer.beeri, sigalit.ifergane\}@gmail.com}
}

\date{} 
\begin{document}

\title{\textit{MLography}: An Automated Quantitative Metallography Model for Impurities Anomaly Detection using Novel Data Mining and Deep Learning Approach}

\maketitle

\begin{abstract}
The micro-structure of most of the engineering alloys contains some inclusions and precipitates, which may affect their properties, therefore it is crucial to characterize them.
In this work we focus on the development of a state-of-the-art artificial intelligence model for Anomaly Detection named \textit{MLography} to automatically quantify the degree of anomaly of impurities in alloys. For this purpose, we introduce several anomaly detection measures: Spatial, Shape and Area anomaly, that successfully detect the most anomalous objects based on their objective, given that the impurities were already labeled. The first two measures quantify the degree of anomaly of each object by how each object is distant and big compared to its neighborhood, and by the abnormally of its own shape respectively. The last measure, combines the former two and highlights the most anomalous regions among all input images, for later (physical) examination. The performance of the model is presented and analyzed based on few representative cases. We stress that although the models presented here were developed for metallography analysis, most of them can be generalized to a wider set of problems in which anomaly detection of geometrical objects is desired. All models as well as the data-set that was created for this work, are publicly available at: \ULurl{https://github.com/matanr/MLography}.

\end{abstract}
\vspace{-0.5cm}

\section{Introduction}

Material science is focused on the correlation between the chemical composition of the material and micro-structure and its properties \cite{sinha2003physical}. The micro-structure of most of the engineering alloys is based on a more or less homogeny matrix that contained some inclusions and precipitates (for shortness, in this paper we will call both of them \textit{inclusions} or \textit{impurities}), which alter its properties such as strength and ductility, heat and electrical transfer. Hence, it is highly important to characterize those inclusions.
This characterization includes the nature of each inclusion (composition, crystallographic structure, size and shape) as well as more general information such as the surface concentration of inclusions (on a cross-section) and their distribution.

While during the development of a new material such characterization can be done comprehensively and deeply, during routine manufacturing usually it is done only once in a while to ensure the manufacture stability and includes a metallographic cross-section characterization \cite{astm20113}. In such routine examination in many cases the soundness of the material is determined by its similarity to previous samples \cite{standarde112}. This similarity is determined by quantitative parameters such as grain size and the surface concentration of inclusions and also by "softer" parameters such as shape and distribution of the inclusions, which are usually based on an "expert opinion" \cite{astm199745}. Therefore, it is of high interest to change this subjective process to more objective and quantitative one, especially as recently the introduction of data mining, machine learning and computer vision techniques to the quantitative metallography field proved to yield excellent results \cite{decost2019high, dimiduk2018perspectives, ramprasad2017machine, mueller2016machine,  decost2015computer}. 

\begin{figure}[h]
\centering
\begin{subfigure}[b]{0.3\linewidth}
\centering
\includegraphics[height=3.3cm, frame]{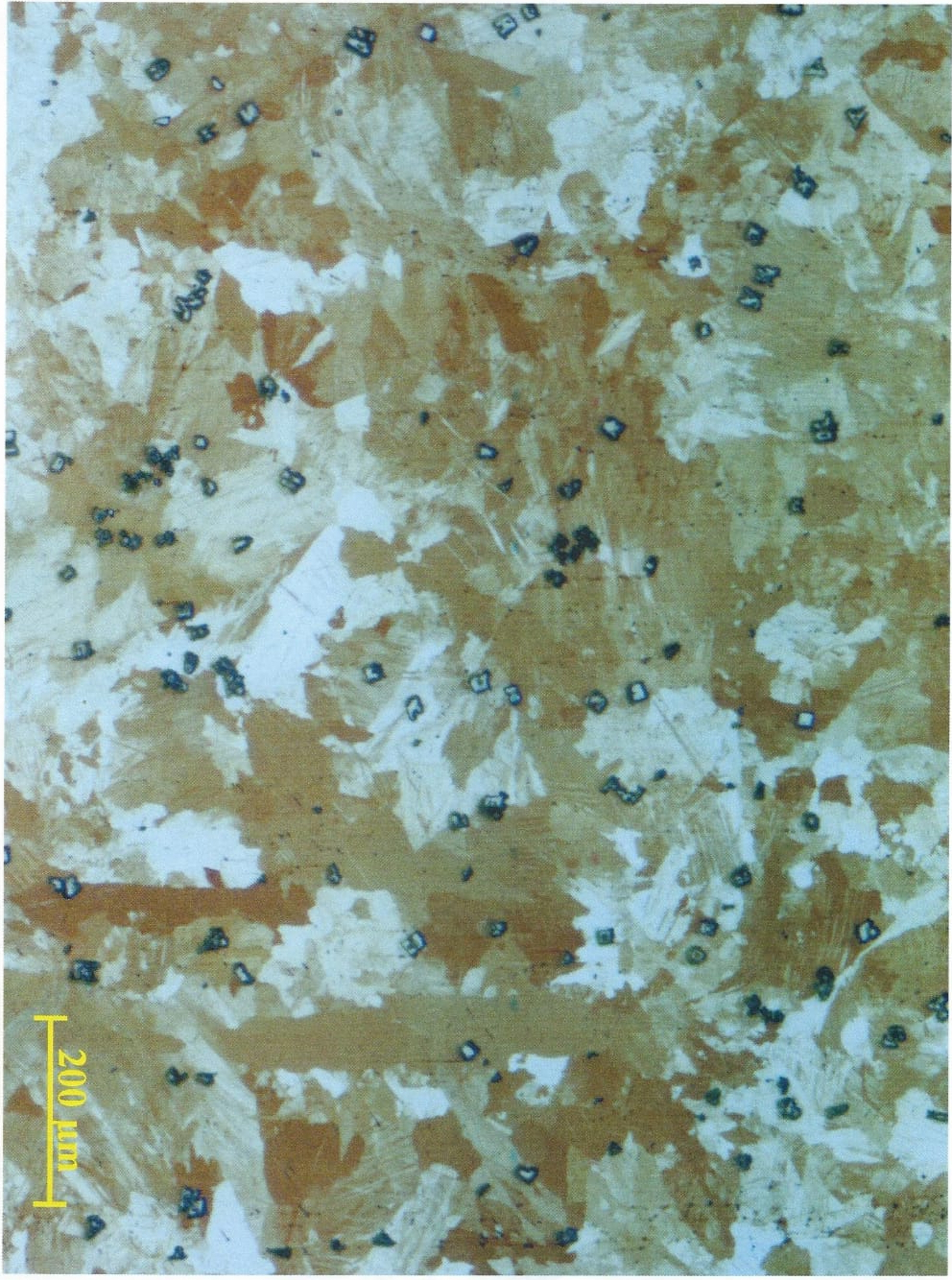}
\caption{Example scan}\label{fig:example_scan}
\end{subfigure}
\begin{subfigure}[b]{.3\linewidth}
\centering
\includegraphics[height=3.3cm, frame]{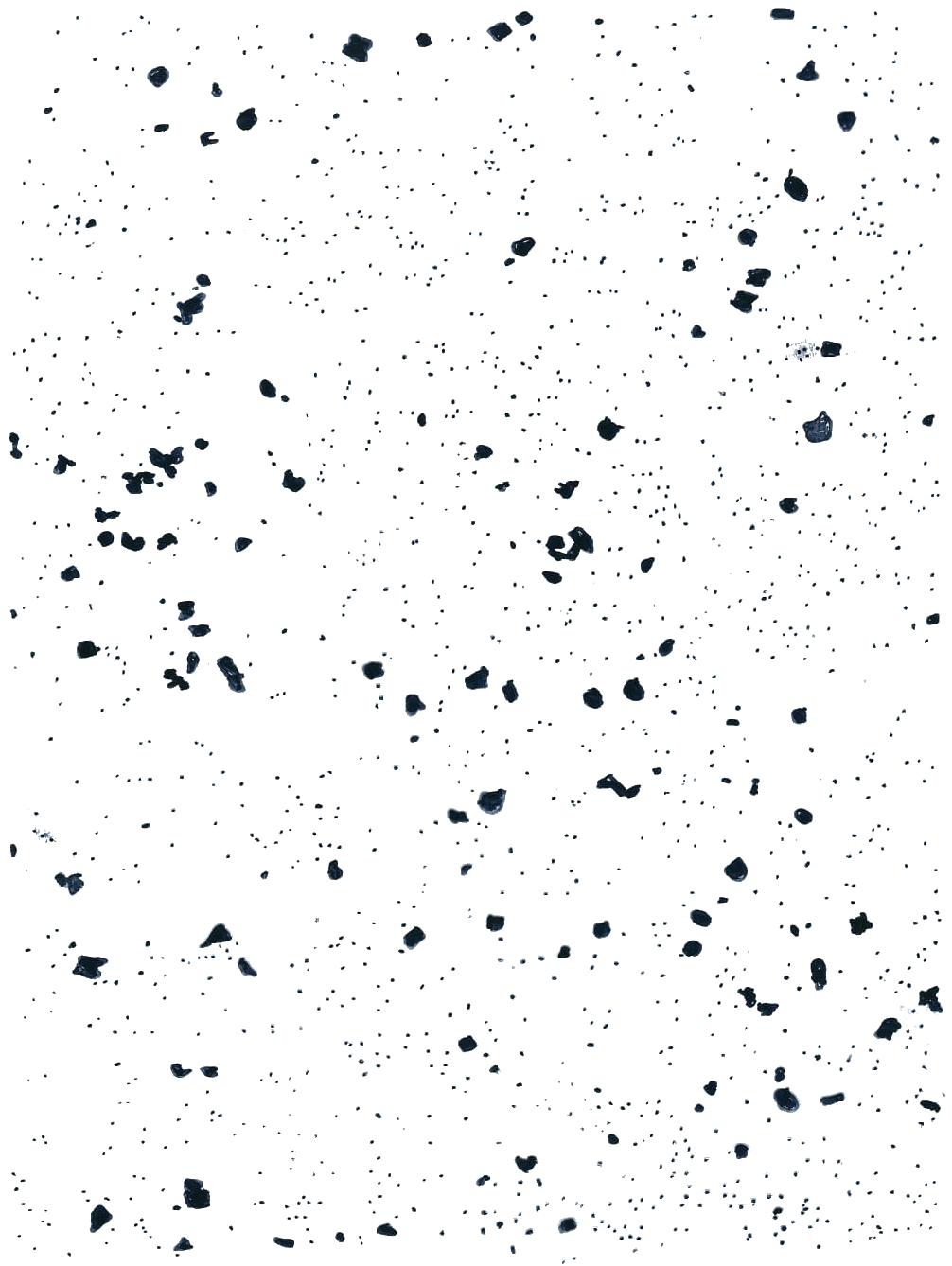}
\caption{Example slide}\label{fig:example_tag}
\end{subfigure}
\begin{subfigure}[b]{.3\linewidth}
\centering
\includegraphics[height=3.3cm, frame]{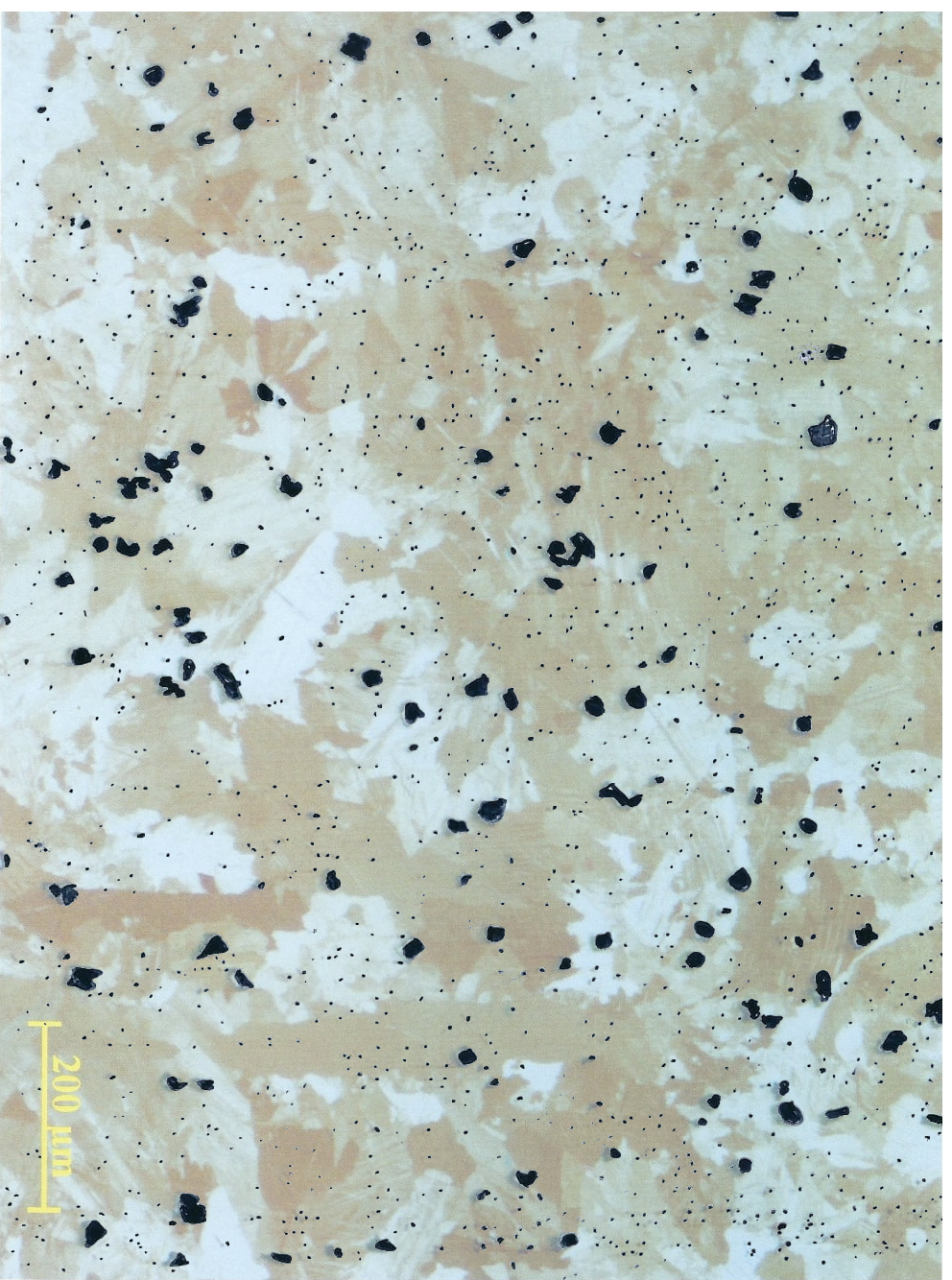}
\caption{Scan with tags}\label{fig:example_scan_tag}
\end{subfigure}
\caption{Metallographic scan and a corresponding tag.}
\end{figure}

In this paper we are trying to do so by a novel data mining and deep learning approach for the U-0.1wt\%Cr alloy which is used as a nuclear fuel.
In this alloy the most abundance inclusion is \textit{Uranium Carbide}, which appears on the 2D metallographic cross-section as dots, spots or long rods, depend on the impurity concentration in the alloy and the thermal profile during casting and cooling to room temperature \cite{nomine1974physical}. For this work, metallographic images (approximately 1.2mm $\times$ 0.8mm, in a scale of 200 $\mu$m) were used (see example in Fig. \ref{fig:example_scan}). A slide was placed on each picture and an expert tagged each inclusion and its boundaries on it (see example in Figures \ref{fig:example_tag}, \ref{fig:example_scan_tag}). As these kind of data-sets are rarely public, except for a few exceptions \cite{decost2017uhcsdb}, we have created a novel data-set with manual tags of impurities over 243 metallographic scans, and we make it publicly available at \cite{MLographyDB}. We present our results on a sample image of tagged impurities from an uranium-chromium alloy scan from the data-set in Fig. \ref{fig:all_plots}.

\section{Anomaly Detection Measures}

\subsection{Spatial Anomaly Measure}\label{chap:spatial_anomaly_measure}

\begin{figure}[]
\begin{center}
    
    \begin{subfigure}[]{0.4\linewidth}
    \centering
    \resizebox{1.\textwidth}{!}{
    \begin{tikzpicture}
        \draw[] (0,0) rectangle (1.25,0.75) node[label={[label distance=0.25cm]60:$i_1$}, pos=.5] {};
        \draw[] (2,0.5) rectangle (3.1,1.25) node[label={[label distance=0.25cm]90:$i_2$}, pos=.5] {};
        \draw[] (-1.2,0.9) rectangle (0.75,2.5) node[label={[label distance=0.85cm]15:$i_3$}, pos=.5] {};
        \draw[] (-0.8, -0.6) rectangle (-0.5, -0.25) node[label={[label distance=0.02cm]180:$i_4$}, pos=.5] {};
        \draw[] (1.1,-0.5) rectangle (2.35,0.15) node[label={[label distance=0.45cm]0:$i_5$}, pos=.5] {};
        
        \node[circle, scale=0.2, draw, fill] (i1) at (0.625,0.375) {};
        \node[circle, scale=0.2, draw, fill] (i2) at (2.55,0.875) {};
        \draw[-, anchor=north] (i1) -- node[above, scale=0.8, sloped] {$d_2$} (i2);
        
        \node[circle, scale=0.2, draw, fill] (i3) at (-0.225,1.7) {};
        \draw[-, anchor=east] (i1) -- node[above, scale=0.8, sloped] {$d_3$} (i3);
        
        \node[circle, scale=0.2, draw, fill] (i4) at (-0.65, -0.425) {};
        \draw[-, anchor=north] (i1) -- node[above left, scale=0.8, sloped] {$d_4$} (i4);
        
        \node[circle, scale=0.2, draw, fill] (i5) at (1.725, -0.175) {};
        \draw[-, anchor=north] (i1) -- node[below right, scale=0.8, sloped] {$d_5$} (i5);
    \end{tikzpicture}
    }
    \caption{Distance function based on fixed point.} \label{fig:fixed_point_dist}
    
    \end{subfigure}
    \hspace{0.2cm}
    \begin{subfigure}[]{0.4\linewidth}
    \centering
    \resizebox{1.\textwidth}{!}{
    \begin{tikzpicture}
        
        \draw[] (0,0) rectangle (1.25,0.75) node[label={[label distance=0.25cm]60:$i_1$}, pos=.5] {};
        \draw[] (2,0.5) rectangle (3.1,1.25) node[label={[label distance=0.25cm]90:$i_2$}, pos=.5] {};
        \draw[] (-1.2,0.9) rectangle (0.75,2.5) node[label={[label distance=0.85cm]15:$i_3$}, pos=.5] {};
        \draw[] (-0.8, -0.6) rectangle (-0.5, -0.25) node[label={[label distance=0.02cm]180:$i_4$}, pos=.5] {};
        \draw[] (1.1,-0.5) rectangle (2.35,0.15) node[label={[label distance=0.45cm]0:$i_5$}, pos=.5] {};
        
        \node[circle, scale=0.2, draw, fill] (i1i2) at (1.25,0.6) {};
        \node[circle, scale=0.2, draw, fill] (i2i1) at (2,0.6) {};
        \draw[-, anchor=north] (i1i2.east) -- node[above, scale=0.8] {$d_2$} (i2i1.west);
        
        \node[circle, scale=0.2, draw, fill] (i1i3) at (0.05,0.75) {};
        \node[circle, scale=0.2, draw, fill] (i3i1) at (0.05,0.9) {};
        \draw[-, anchor=east] (i1i3.north) -- node[left, scale=0.8] {$d_3$} (i3i1.south);
        
        \node[circle, scale=0.2, draw, fill] (i1i4) at (0,0) {};
        \node[circle, scale=0.2, draw, fill] (i4i1) at (-0.5, -0.25) {};
        \draw[-, anchor=north] (i1i4) -- node[above, scale=0.8, sloped] {$d_4$} (i4i1);
    \end{tikzpicture}
    }
    \caption{Suggested distance function.} \label{fig:imp_dist}
    \end{subfigure}
 
\end{center}
\caption{Possible distance functions for geometric objects.} \label{fig:two_dist_funcs}
\end{figure}
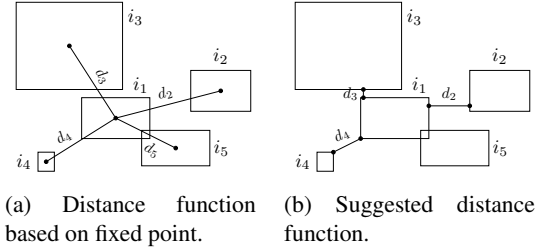

Unsupervised Distance-based is one of the most common setups for anomaly detection \cite{goldstein2016comparative}. In this approach, an object is considered as an \textit{outlier} based on its spatial properties, most common among those properties is how distant the object is from its neighbourhood. A unified distance-based notion of anomaly presented in \cite{knox1998algorithms}:
\textit{An object $O$ in data-set $T$ is a $DB(p,D)-outlier$ if at least fraction $p$ of the objects in $T$ are $\geq$ distance $D$ from $O$}, where $DB$ stands for Distance-Based.
Although this notion is applicable for generalizing statistical anomaly detection in distributions such as Normal, Exponential and Poisson distributions, it lacks few important properties: It is not able to produce scores of anomaly; It requires the user to provide the distance $D$; And most importantly it does not treat objects with shapes of a positive area, as the impurities in our study. 
Another common distance-based anomaly detection approach, K$^{th}$-Nearest-Neighbour \cite{ramaswamy2000efficient, zhang2009new} henceforth K$^{th}$-NN, defines outliers by their distance from their $k^{th}$ nearest neighbour, and sorts them by that measure. Indeed, this approach allows one to order object by a measure that indicates how that object is \textit{distant} from its neighbourhood. K$^{th}$-NN was compared to other 18 different unsupervised anomaly detection algorithms, on 10 data-sets and it was found to outperform all other algorithms with regard to accuracy, determinism and the ability to detect global anomaly \cite{goldstein2016comparative}. However, in this study we focus on anomaly detection for geometric objects with a positive area with high emphasis on their size, i.e. the desired spatial measure should consider the areas of the impurities in order to score each impurity by how it is \textit{distant} and \textit{big} compared to its neighbourhood. To that end, we present a novel approach for \textit{spatial anomaly detection for positive area geometric objects}. Our spatial anomaly detection approach first defines a pseudo-semi-metric distance function between two geometric objects by the distance between their \textit{Straight Bounding Rectangles}. We use rectangles since they are simplistic and computationally easy to calculate for each impurity, yet accurate enough - both in terms of Contour Approximation as well as edges distance calculation. We note that a trivial approach for implementing this distance function might be using the euclidean distance between any two points on the objects, e.g. the centers of the objects as presented in Fig. \ref{fig:fixed_point_dist}. But in case of almost-intersecting two big objects, this approach will yield a much higher distance than what is expected, since their borders are much closer than their centers (e.g. the distance between $i_1$ and $i_3$ in Fig. \ref{fig:fixed_point_dist}). A similar argument can be made on any fixed points residing on the objects.
Our function is summarized with the following 4 representative cases in Fig. \ref{fig:imp_dist}.
The distance between an object $i_1$ to another object, in case of non-intersecting rectangles ($i_2,i_3,i_4$), is defined by the shortest euclidean distance between the boundaries of the two enclosing rectangles (i.e. distance between the closest two edges in the first two cases, and the distance between the closest corner vertices in the last case respectively). The distance between two intersecting enclosing rectangles ($i_1, i_5$) is simply defined as 0. It can be shown that this distance measure satisfies the symmetry axiom, and that for each two objects {$o_1, o_2$} the distance is $d(o_1,o_2) \geq 0$ but the triangle inequality axiom is not met and not necessarily $d(o_1,o_2) = 0$ means that $o_1=o_2$.

Next we present a modified version of the classical K$^{th}$-NN algorithm \cite{cover1967nearest}: \textit{Weighted}-K$^{th}$-Nearest-Neighbor henceforth WK$^{th}$-NN, in which each object $i$ refers to the distance between $i$ to its neighbourhood (defined above), along with the proportion between $i$'s area to its neighbours. This allows having the spatial anomaly score to be calculated as a function of how the object $i$ is \textit{distant} from its neighbours, and also as how it is \textit{big} compared to its neighbours. We now describe the algorithm, which is summarized in Algorithm \ref{WKthNN}. As in K$^{th}$-NN, the algorithm is parameterized by \textit{k} - a constant that states how far is the neighbour from which we calculate the distance from.

\begin{algorithm}
\caption{Weighted-K$^{th}$NN}\label{WKthNN}
\begin{algorithmic}[1]
\Procedure{WeightedDist}{$i, o$}
  \State \textbf{return} $\left(\tfrac{\mbox{\sc Area}(i)}{\mbox{\sc Area}(o)}\right) ^ {c_1} * \mbox{\sc ImpurityDist}(i,o)$ \label{s_alg:calc_w_dist}
 \EndProcedure
 \newline
\Procedure{WK$^{th}$NN}{$k$}
\begin{flushleft}
	$SS$: list of size $|\mathcal{I}|$ \Comment{$SS$ as $SpatialScores$}
\end{flushleft}
    \For{$i\in \mathcal{I}$} \label{s_alg:all_imps}
        \State $l = [\mbox{\sc WeightedDist}(i,o); \forall o \in \mathcal{I} \setminus \{i\}]$ \label{s_alg:calc_others}
        \State $\mbox{\sc Sort}(l)$ \label{s_alg:sort}
        \State $SS_i = \mbox{\sc Area}(i) * l[k]^{c_2}$ \label{s_alg:self_area}
  \EndFor
  \State $\forall ss \in SS, \frac{ss - min(SS)}{max(SS) - min(SS)}$ \label{s_alg_norm}\Comment{Min-Max norm}
  \State \textbf{return} $SS$ 
 \EndProcedure
\end{algorithmic}
\end{algorithm}

The procedure $\mbox{\sc WeightedDist}$ calculates the weighted distance measure between the impurity $i$ and the other impurity $o$. This measure combines the proportion between the areas of $i$ and $o$ and the distance between them (defined in Fig. \ref{fig:imp_dist}). The main algorithm, \mbox{\sc W$K^{th}$NN}, iterates over all objects (impurities in our case), $\mathcal{I}$, in line \ref{s_alg:all_imps}. For each object $i$, it calculates for all other objects $o$, the weighted distance in line \ref{s_alg:calc_others}. Then, it sorts the returned distances in line \ref{s_alg:sort}, and adds a factor of $\mbox{\sc Area}(i)$ in line \ref{s_alg:self_area}, in order to emphasize the significance of that object. Finally, when the iteration over all objects completes, we normalize and save the spatial scores of each object $i$ in $SS_i$. The constants $c_1, c_2$ were set to $4,2$ respectively, but we encourage users to determine the values of the constants $c_1, c_2$, to suit best to their data-sets. The output of the spatial anomaly detection algorithm on the input image with $k=50$ is presented in Fig. \ref{fig:spatial}.

\subsection{Shape Anomaly Measure}

Another important property of each geometric object is its shape, or how "\textit{close to some objective shape}" is it, which in our case, how symmetric and how close the impurity to a circle.
Examining the output of the spatial anomaly detection algorithm may give the idea that spatial anomaly detection is sufficient for describing the degree of anomaly in each object, since it successfully marks objects that are clear to be anomalous with high anomaly score.
However, the spatial anomaly measure is not able to distinguish between an object that is not that big and distant compared to its neighbourhood and does not have \textit{anomalous} shape (e.g. an 'O' shape impurity), with an object of the same distance and size compared to its neighbourhood, but with a much more anomalous shape (e.g. an 'X' shape impurity). Therefore, a consideration of the actual shape of each impurity is necessary to determine whether it is an outlier or not.
A trivial measure for non-symmetric shape anomaly detection might be finding for each object $i$ its smallest enclosing circle object, $c$ (or some other basic geometric shape as in \cite{igathinathane2008shape}) and setting $i$'s shape anomaly score as:
\begin{equation}
\tfrac{\mbox{\sc Area}(c) - \mbox{\sc Area}(i)}{\mbox{\sc Area}(c)} \label{eq_circle_diff}
\end{equation} 
This measure indeed catches the most anomalous and non-anomalous objects based on their shape (i.e. impurities of a shape with area far smaller than their smallest enclosing circle's area, and impurities of a shape very close to a circle respectively), but it fails to classify properly objects in the middle of the scale, as can be seen in Fig. \ref{fig:circle_diff}.

 \begin{figure}[htb]
\begin{center}
  \includegraphics[width=.6\linewidth]{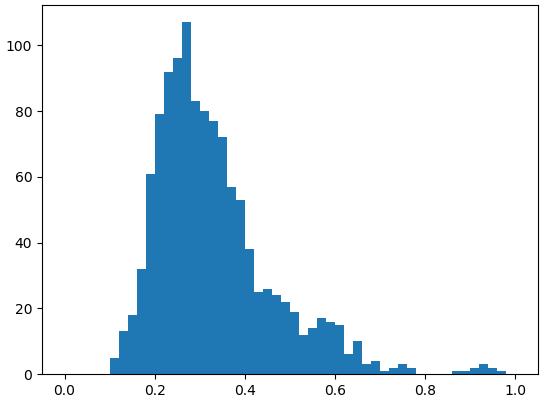}
\end{center}
   \caption{Histogram of circle difference scores in Fig. \ref{fig:circle_diff}} \label{fig:circle_hist}
\end{figure}

 For example, impurities of the anomalous shape 'X' are marked only in the middle of the scale (e.g. the impurity within the black rectangle in Fig. \ref{fig:circle_diff}), together with not-that-anomalous ellipse-shaped impurities, as their shape's area is not that far from their smallest enclosing circle's area, although they should have appeared higher in the shape anomaly score scale. Indeed, Fig. \ref{fig:circle_hist} shows that there is a decent separation between the most anomalous impurities (scores $\geq 0.6$ ) and the rest of the impurities (scores around $0.3$), but the right tail of the distribution is quite long, which imposes noise to the model.
 Thus, from the non-linear nature of the problem at hand, we turned to train a Deep Convolutional Auto-Encoder Neural Network henceforth AE, for shape anomaly detection (also called Replicator Neural Networks) \cite{hawkins2002outlier, dau2014anomaly} which enforces the network to reconstruct images that are similar to the images from the training set, and, hopefully, to fail reconstructing images that are not similar to the images in the training set.
 
 As we already stated, the circle difference measure from Equation \ref{eq_circle_diff} is a good estimate for shape anomaly in the two ends of the scale - the most anomalous impurities and most non-anomalous impurities, which we will denote normal impurities from now on. Thus, one can train an AE network in an \textit{unsupervised} manner by providing the network, in its training phase, couples of all normal impurities as the training samples and copies of themselves as their labels. That is, fix a threshold for normal impurities and take all impurities with anomaly score lower than that threshold.
 \pgfplotsset{every tick label/.append style={font=\big}}

\begin{figure}[hb]
\centering
\resizebox{0.45\textwidth}{!}{
\begin{tikzpicture}
\tikzstyle{connection}=[ultra thick,every node/.style={sloped,allow upside down},draw=\edgecolor,opacity=0.7]
\tikzstyle{copyconnection}=[ultra thick,every node/.style={sloped,allow upside down},draw={rgb:blue,4;red,1;green,1;black,3},opacity=0.7]

\node[canvas is zy plane at x=0] (temp) at (-3,0,0) {\includegraphics[width=8cm,height=8cm, frame]{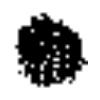}};

\pic[shift={(0,0,0)}] at (0,0,0) 
    {Box={
        name=conv1,
        caption= ,
        xlabel={{256, }},
        zlabel=100,
        fill=\ConvColor,
        height=40,
        width=3,
        depth=40,
        }
    };

\pic[shift={ (0,0,0) }] at (conv1-east) 
    {Box={
        name=pool1,
        caption= ,
        fill=\PoolColor,
        opacity=0.5,
        height=25,
        width=1,
        depth=25
        }
    };

\pic[shift={(1,0,0)}] at (pool1-east) 
    {Box={
        name=conv2,
        caption= ,
        xlabel={{256, }},
        zlabel=50,
        fill=\ConvColor,
        height=25,
        width=3,
        depth=25
        }
    };

\draw [connection]  (pool1-east)    -- node {\midarrow} (conv2-west);

\pic[shift={ (0,0,0) }] at (conv2-east) 
    {Box={
        name=pool2,
        caption= ,
        fill=\PoolColor,
        opacity=0.5,
        height=13,
        width=1,
        depth=13
        }
    };

\pic[shift={(1,0,0)}] at (pool2-east) 
    {Box={
        name=conv3,
        caption= ,
        xlabel={{256, }},
        zlabel=25,
        fill=\ConvColor,
        height=13,
        width=3,
        depth=13
        }
    };

\draw [connection]  (pool2-east)    -- node {\midarrow} (conv3-west);

\pic[shift={ (0,0,0) }] at (conv3-east) 
    {Box={
        name=pool3,
        caption= ,
        fill=\PoolColor,
        opacity=0.5,
        height=7,
        width=1,
        depth=7
        }
    };

\pic[shift={(1,0,0)}] at (pool3-east) 
    {Box={
        name=conv4,
        caption= ,
        xlabel={{256, }},
        zlabel=13,
        fill=\ConvColor,
        height=7,
        width=3,
        depth=7
        }
    };

\draw [connection]  (pool3-east)    -- node {\midarrow} (conv4-west);

\pic[shift={ (0,0,0) }] at (conv4-east) 
    {Box={
        name=pool4,
        caption= ,
        fill=\PoolColor,
        opacity=0.5,
        height=3,
        width=1,
        depth=3
        }
    };

\pic[shift={(1,0,0)}] at (pool4-east) 
    {Box={
        name=conv5,
        caption= ,
        xlabel={{256, }},
        zlabel=7,
        fill=\ConvColor,
        height=3,
        width=3,
        depth=3
        }
    };

\draw [connection]  (pool4-east)    -- node {\midarrow} (conv5-west);

\pic[shift={ (0,0,0) }] at (conv5-east) 
    {Box={
        name=unpool1,
        caption= ,
        fill=\UnpoolColor,
        opacity=0.5,
        height=7,
        width=1,
        depth=7
        }
    };

\pic[shift={(1,0,0)}] at (unpool1-east) 
    {Box={
        name=conv6,
        caption= ,
        xlabel={{256, }},
        zlabel=10,
        fill=\ConvColor,
        height=5,
        width=3,
        depth=5
        }
    };

\draw [connection]  (unpool1-east)    -- node {\midarrow} (conv6-west);

\pic[shift={ (0,0,0) }] at (conv6-east) 
    {Box={
        name=unpool2,
        caption= ,
        fill=\UnpoolColor,
        opacity=0.5,
        height=10,
        width=1,
        depth=10
        }
    };

\pic[shift={(1,0,0)}] at (unpool2-east) 
    {Box={
        name=conv7,
        caption= ,
        xlabel={{256, }},
        zlabel=16,
        fill=\ConvColor,
        height=8,
        width=3,
        depth=8
        }
    };

\draw [connection]  (unpool2-east)    -- node {\midarrow} (conv7-west);

\pic[shift={ (0,0,0) }] at (conv7-east) 
    {Box={
        name=unpool3,
        caption= ,
        fill=\UnpoolColor,
        opacity=0.5,
        height=16,
        width=1,
        depth=16
        }
    };

\pic[shift={(1,0,0)}] at (unpool3-east) 
    {Box={
        name=conv8,
        caption= ,
        xlabel={{256, }},
        zlabel=28,
        fill=\ConvColor,
        height=14,
        width=3,
        depth=14
        }
    };

\draw [connection]  (unpool3-east)    -- node {\midarrow} (conv8-west);

\pic[shift={ (0,0,0) }] at (conv8-east) 
    {Box={
        name=unpool4,
        caption= ,
        fill=\UnpoolColor,
        opacity=0.5,
        height=28,
        width=1,
        depth=28
        }
    };

\pic[shift={(1,0,0)}] at (unpool4-east) 
    {Box={
        name=soft1,
        caption= ,
        xlabel={{" ","dummy"}},
        zlabel=500,
        fill=\SoftmaxColor,
        opacity=0.8,
        height=3,
        width=1.5,
        depth=10
        }
    };

\draw [connection]  (unpool4-east)    -- node {\midarrow} (soft1-west);

\pic[shift={(1,0,0)}] at (soft1-east) 
    {Box={
        name=soft2,
        caption= ,
        xlabel={{" ","dummy"}},
        zlabel=10000,
        fill=\SoftmaxColor,
        opacity=0.8,
        height=3,
        width=1.5,
        depth=30
        }
    };

\draw [connection]  (soft1-east)    -- node {\midarrow} (soft2-west);

\node[canvas is zy plane at x=0] (temp) at (18,0,0) {\includegraphics[width=8cm,height=8cm,frame]{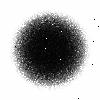}};

\draw[dashed, ultra thick,every node/.style={sloped,allow upside down},draw=blue,opacity=0.7] (16,0,0) --  (18,0,0);

\draw[dashed, ultra thick,every node/.style={sloped,allow upside down},draw=blue,opacity=0.7] (-3,0,0) --  (0,0,0);

\end{tikzpicture}
}
\caption{The AE network architecture. We used different paddings for the convolutional layers in the Decoder, in order to reduce memory footprint and fit the model into 32GB Tesla V100 GPU \cite{negevhpc}, using TensorFlow \cite{abadi2016tensorflow}. Yellow layers are Convolutional layers (5x5), Orange are Max-Pooling layers, Blue are Up-Sampling layers, Purple are Fully Connected layers and the left-most and right-most images are example input and output images respectively.}
\label{AE_layers}
\end{figure}
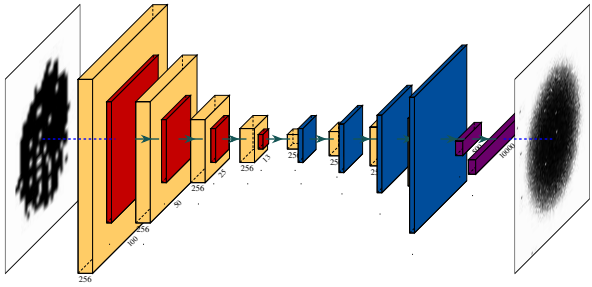
\begin{table*}[!b]
  \centering
  \resizebox{0.85\textwidth}{!}{
  \begin{tabular}{ | c | c || c | c | c || c | c | c |}
    \hline
    Name & Image & \begin{tabular} {l} Model 1 \\ Recon' \end{tabular} & \begin{tabular} {l} Model 1 \\ Post-Recon' \end{tabular} & \begin{tabular} {l} Model 1 \\ MSE \end{tabular} &  \begin{tabular} {l} Model 2 \\ Recon' \end{tabular} & \begin{tabular} {l} Model 2 \\ Post-Recon' \end{tabular} & \begin{tabular} {l} Model 2 \\ MSE \end{tabular} \\ \hline \hline 
    \begin{minipage}{.09\linewidth}
    \centering
      $Norm_1$
    \end{minipage}
    &
    \begin{minipage}{.09\linewidth}
      \includegraphics[width=\linewidth]{shape_blank_label/normal.jpg}
    \end{minipage}
    &
    \begin{minipage}{.09\linewidth}
      \includegraphics[width=\linewidth]{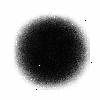}
    \end{minipage}
    & 
    \begin{minipage}{.09\linewidth}
      \includegraphics[width=\linewidth]{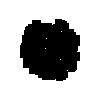}
    \end{minipage}
    &
    \begin{minipage}{.09\linewidth}
    \centering
      $45,479$
    \end{minipage}
    &
    \begin{minipage}{.09\linewidth}
      \includegraphics[width=\linewidth]{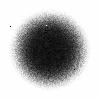}
    \end{minipage}
    & 
    \begin{minipage}{.09\linewidth}
      \includegraphics[width=\linewidth]{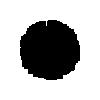}
    \end{minipage}
    &
    \begin{minipage}{.09\linewidth}
    \centering
      $45,919$
    \end{minipage}
    \\ \hline
    \hline
    \begin{minipage}{.09\linewidth}
    \centering
      $Norm_2$
    \end{minipage}
    &
    \begin{minipage}{.09\linewidth}
      \includegraphics[width=\linewidth]{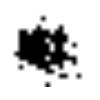}
    \end{minipage}
    &
    \begin{minipage}{.09\linewidth}
      \includegraphics[width=\linewidth]{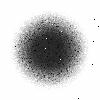}
    \end{minipage}
    & 
    \begin{minipage}{.09\linewidth}
      \includegraphics[width=\linewidth]{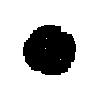}
    \end{minipage}
    &
    \begin{minipage}{.09\linewidth}
    \centering
      $51,233$
    \end{minipage}
    &
    \begin{minipage}{.09\linewidth}
      \includegraphics[width=\linewidth]{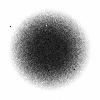}
    \end{minipage}
    & 
    \begin{minipage}{.09\linewidth}
      \includegraphics[width=\linewidth]{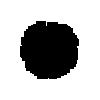}
    \end{minipage}
    &
    \begin{minipage}{.09\linewidth}
    \centering
      $46,668$
    \end{minipage}
    \\ \hline
    \hline
    \begin{minipage}{.09\linewidth}
    \centering
      $Anom_1$
    \end{minipage}
    &
    \begin{minipage}{.09\linewidth}
      \includegraphics[width=\linewidth]{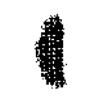}
    \end{minipage}
    &
    \begin{minipage}{.09\linewidth}
      \includegraphics[width=\linewidth]{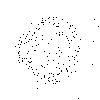}
    \end{minipage}
    & 
    \begin{minipage}{.09\linewidth}
      \includegraphics[width=\linewidth]{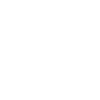}
    \end{minipage}
    &
    \begin{minipage}{.09\linewidth}
    \centering
      $64,585$
    \end{minipage}
    &
    \begin{minipage}{.09\linewidth}
      \includegraphics[width=\linewidth]{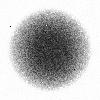}
    \end{minipage}
    & 
    \begin{minipage}{.09\linewidth}
      \includegraphics[width=\linewidth]{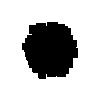}
    \end{minipage}
    &
    \begin{minipage}{.09\linewidth}
    \centering
      $48,869$
    \end{minipage}
    \\ \hline
    \hline
    \begin{minipage}{.09\linewidth}
    \centering
      $Anom_2$
    \end{minipage}
    &
    \begin{minipage}{.09\linewidth}
      \includegraphics[width=\linewidth]{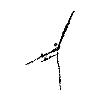}
    \end{minipage}
    &
    \begin{minipage}{.09\linewidth}
      \includegraphics[width=\linewidth]{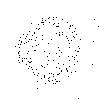}
    \end{minipage}
    & 
    \begin{minipage}{.09\linewidth}
      \includegraphics[width=\linewidth]{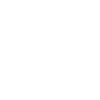}
    \end{minipage}
    &
    \begin{minipage}{.09\linewidth}
    \centering
      $64,527$
    \end{minipage}
    &
    \begin{minipage}{.09\linewidth}
      \includegraphics[width=\linewidth]{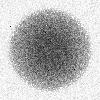}
    \end{minipage}
    & 
    \begin{minipage}{.09\linewidth}
      \includegraphics[width=\linewidth]{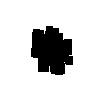}
    \end{minipage}
    &
    \begin{minipage}{.09\linewidth}
    \centering
      $54,617$
    \end{minipage}
    \\ \hline
  \end{tabular}
  }
\caption{Reconstruction of impurities in both AE models. Model 1 was trained using same images as labels for normal impurities, and blank images as labels for anomalous impurities. Model 2 was trained only with normal impurities.}
\label{fig:recon}
\end{table*}
 We present in this work a novel approach to empower the separation capability of AE networks, in which together with the normal couples of input and label images, the network is provided with couples of all the most anomalous images as input and blank images as labels. This will urge the network to reconstruct successfully the normal images, and to return a noisy-blank image upon an anomalous input, or in our case, anomalous impurity. This in turn, will yield in higher reconstruction loss for anomalous impurities, therefore normalizing the reconstruction loss and using it as shape anomaly measure will offer a higher separation between normal and abnormal impurities. We stress that one big advantage of using neural networks is that it requires no assumption about the data, therefore one can employ the presented technique on any predefined 'normal' and 'abnormal' objects (in our case, difference from a circle).
We set the threshold for normal impurities with $0.3$, and for anomalous impurities with $0.55$, and normalized and scaled all input images into the same size of $100 \times 100$ pixels. We note that albeit the size feature is not preserved in this measure, we still consider it in the spatial anomaly measure in chapter \ref{chap:spatial_anomaly_measure}. Then we trained an AE network of the architecture presented in Fig. \ref{AE_layers}. The achieved reconstruction results on several use-cases, consisting of two normal-shaped impurities, $Norm_1, Norm_2$, and two anomalous-shaped impurities, $Anom_1, Anom_2$, are presented in Table \ref{fig:recon}.

 Column \textit{Image} holds the representative image of each impurity and \textit{Model 1 Recon'} holds the reconstructed image from the AE model trained on both normal and anomalous impurities. As we can see, there is a strong separation between the normal and anomalous impurities' reconstruction in the first model, as in the first two impurities the reconstruction is a well-formed circle (with a varying intensity with respect to the degree of anomaly), and for the last ones the reconstruction is a noisy-circle. For even stronger separation, we applied post-processing (threshold, erode-dilate) on the output of the AE which is shown in \textit{Model 1 Post-Recon'} column, and by that obtaining circles of different sizes for each of the normal impurities, and a blank image for the anomalous impurities. The column \textit{Model 1 MSE} shows the Mean Squared Error (MSE) as the reconstruction loss, between the input image and the reconstructed image after post-processing. As we can clearly see, the first impurity is the most 'normal' impurity and the last two impurities are much more anomalous.
 
 Conversely, the reconstruction results of the same input impurities, on an AE trained on only normal set of impurities are presented in the columns \textit{Model 2 Recon', Model 2 Post-Recon'} and \textit{Model 2 MSE}.

 As we can see, the intensity of the reconstructed circle scales negatively with the degree of anomaly, thus again yielding circles of different sizes in the post-processed reconstructions. Additionally, the MSE of the most anomalous impurity, $Anom_2$, is significantly higher than of the most symmetric impurity, $Norm_1$, but the difference between the errors of $Norm_2$ and $Anom_1$ is mild, thus the separation between the normal and anomalous impurities is flawed. 
  Figures \ref{fig:shape_blank}, \ref{fig:shape_same} present the output of both models, in each the normalized reconstruction losses serves as the shape anomaly measure.

The model that utilizes blank images as labels for anomalous impurities in the training phase greatly outperforms the second, since it has a more acute separation between normal-shaped and anomalous-shaped impurities, and it marks the anomalous 'X'-shaped impurities with high anomaly score. We therefore use this model.
The previously purposed spatial anomaly measure - combined by simple multiplication and normalization with the shape anomaly measure - is presented in Fig. \ref{fig:shape_spatial}. This measure extremely reduces the noise we had in the spatial anomaly measure, while emphasizing the degree of anomaly of impurities that are anomalous based on their shape and compared to their neighbourhood.

\subsection{Area Anomaly Measure}
As previously explained, an important application for anomaly detection in the context of materials sciences and others, is detecting defects. This defects normally span an anomalous \textit{area} of several objects, rather that just a single anomalous object \cite{deepak2016anomaly}. For this reason we present a novel clustering algorithm, which we call \textit{Market Clustering}, that divides impurities into anomalous areas, based on the anomaly scores of the impurities from the previous anomaly measures. The name of the algorithm is inspired from the 'purchasing power' of each area/cluster and the economic decisions that it should take in order to grow and merge with other big clusters. In fact, the size, reach and anomaly score of each cluster is determined based on the anomaly score of the objects from the previous measures.
The returned clusters are then ranked based on a measure that we later describe, and the anomalous areas beyond some pre-determined threshold are suggested for further physical tests.
We next present the algorithm in Algorithm \ref{mc} and then describe its actions.

\begin{algorithm}
\caption{Market-Clustering}\label{mc}
\begin{algorithmic}[1]
\Procedure{MarketClustering}{$k, scores$}
\begin{flushleft}
	$\mathcal{A}$: for $i\in \mathcal{I}$ that participated in some auction, stores the highest bid for $i$ from some cluster
\end{flushleft}
    \State $clusters = \mbox{\sc InitClusters}(k, scores)$ \label{a_alg:init_clusters}
    \State $status =$ not converged
    \While{$status \neq$ converged} \label{a_alg:mc_while}
        \State $status =$ converged
        \State sort $clusters$ by their $wallet$
        \For{$c \in clusters$} \label{a_alg:mc_for}
            \State $(i,o) \leftarrow$ cheapest couple; $i\in c.\mathcal{I}$, $o \notin c.\mathcal{I}$ 
            \myindent{2.8}based on \mbox{\sc Price}($i,o$) \label{a_alg:find_cheapest_couple}
            \If{$o \in \mathcal{A}$ and $c.\mathcal{W} \leq \mathcal{A}[o]$} \label{a_alg:higher_bid}
                \State \textbf{goto} \ref{a_alg:find_cheapest_couple} \Comment{there is higher bid}
            \EndIf
            \State $status = \mbox{\sc AttemptToExpand}(c, i, o, \mathcal{A})$ \label{a_alg:attempt_to_expand}
            \If{$status = $ merged} \label{a_alg:if_merged}
                \State \textbf{goto} \ref{a_alg:mc_while}
            \EndIf
      \EndFor
  \EndWhile
  \State \textbf{return} $clusters$ 
 \EndProcedure
 \newline
\Procedure{InitClusters}{$k, scores$} \label{a_alg:init_clusters_start}
  \For{$i\in [1,k]$}
    \State $core$ = imp' with the $i$-highest anomaly score
    \State $c$:  New Cluster
    \State $c.\mathcal{C}= [coreImpurity]$   \Comment{core impurities} \label{a_alg:core}
    \State $c.\mathcal{I} = [coreImpurity]$   \Comment{impurities inside} \label{a_alg:inside}
    \State $c.\mathcal{W} = \mbox{\sc F} \left(scores[coreImpurity]\right)$ \Comment{wallet} \label{a_alg:wallet}
    \State $clusters.$\mbox{\sc Append}$\left(c\right)$
  \EndFor
  \State \textbf{return} $clusters$ \label{a_alg:init_clusters_end}
\EndProcedure
\newline
\Procedure{AttemptToExpand}{$c, i, o, \mathcal{A}$}
\If{$\exists c' \in clusters$ s.t. $o \in c'.\mathcal{C}$} \label{a_alg:other_is_core}
    \State $\mathcal{A}[o] = c.\mathcal{W}$ \Comment{place bid}
    \State $c.\mathcal{W} = c.\mathcal{W} + c'.\mathcal{W}$ \Comment{combine wallets}
    \State $c.\mathcal{C} = \mbox{\sc Concatenate}(c.\mathcal{C}, c'.\mathcal{C})$
    \State $c.\mathcal{I} = \mbox{\sc Concatenate}(c.\mathcal{I}, c'.\mathcal{I})$
    \State $clusters.\mbox{\sc Remove}(c')$
    \State \textbf{return} merged
\ElsIf{$c.\mathcal{W} \geq \mbox{\sc Price}(i,o)$} \label{a_alg:there_is_money}
    \State $\mathcal{A}[o] = c.\mathcal{W}$ \Comment{place bid}
    \State $c.\mathcal{W} = c.\mathcal{W} - \mbox{\sc Price}(i,o)$ \Comment{pay}
    \State $c.\mathcal{I}.\mbox{\sc Append}(o)$
    \If{$\exists c' \in clusters$ s.t. $o \in c'.\mathcal{I}$}
        \State $c'.\mathcal{I}.\mbox{\sc Remove}(o)$
    \EndIf
\EndIf
\State \textbf{return} converged
\EndProcedure
\end{algorithmic}
\end{algorithm}

The procedure \mbox{\sc MarketClustering} is the main procedure of the algorithm. It receives as input a parameter $k$ - number of initial clusters, and a list $scores$ - impurities' anomaly scores, which in our case are the combination of the spatial and shape measures. First we initialize the list $clusters$ in line \ref{a_alg:init_clusters}, by defining in \mbox{\sc InitClusters} for each cluster $c$ its \textit{Core impurities} list - $c.\mathcal{C}$, \textit{Impurities inside} list - $c.\mathcal{I}$, and \textit{Wallet balance} variable - $c.\mathcal{W}$. 
The procedure in lines \ref{a_alg:init_clusters_start} - \ref{a_alg:init_clusters_end} does that by setting the core impurities of $k$ clusters with the $k$ most anomalous impurities (lines \ref{a_alg:core}, \ref{a_alg:inside}).
Core impurities represent each cluster, thus the degree of anomaly of each cluster is first defined by the anomaly score of the initial core impurity and it is stored in the wallet of the cluster in line \ref{a_alg:wallet}.
The loop in line \ref{a_alg:mc_while} iterates until convergence, and in each iteration it sorts the clusters by their wallet and initiates the main loop in line \ref{a_alg:mc_for}. This loop iterates over all clusters, and for each cluster $c$ it finds in line \ref{a_alg:find_cheapest_couple} the cheapest couple $(i,o)$, s.t. $i \in c.\mathcal{I}$ is an impurity inside of it, and $o\in \mathcal{I}\setminus c.\mathcal{I}$ is an impurity outside of it, under some parametric price function \mbox{\sc Price}($i,o$). This couple in fact implies that the cheapest impurity for cluster $c$ to append is $o$, and the price for it is \mbox{\sc Price}($i,o$). We suggest that the price will be a function of distance and anomaly score, i.e. the price should decrease as the distance decreases to encourage clusters to be continuous, and as the anomaly scores increase to instruct clusters to expand towards anomalous impurities and cover a larger anomalous area. We later present our parametric price function in Algorithm \ref{a_alg:price}.
In line \ref{a_alg:higher_bid} we make use of $\mathcal{A}$ which is a list that stores for each impurity $i\in \mathcal{I}$ what is the cluster with the highest wallet balance that tried to append $i$ to himself. In order to prevent clusters from fighting and emptying their wallets over impurities, we allow $c$ to proceed to line \ref{a_alg:attempt_to_expand} only if it is the cluster with the highest balance that has attempted to append $o$ until now. In line \ref{a_alg:attempt_to_expand}, $c$ attempts to expand its reach by calling the procedure \mbox{\sc AttemptToExpand}. In line \ref{a_alg:other_is_core} we check if the other impurity $o$ is a core impurity of some other cluster $c'$. If it is, first $\mathcal{A}$ is updated with the new bid on $o$, and then the clusters $\{c,c'\}$ are merged into $c$. Otherwise, as an utilization of \textit{credit with no overdraft} policy, if there is enough credit in the wallet of $c$, again $\mathcal{A}$ is updated with the new bid on $o$, and $c$ pays for and appends $o$. Then we check in line \ref{a_alg:if_merged} if a merge has occurred, and if it did, we sort the clusters again by their wallet balance and proceed to iterate over all new clusters in line \ref{a_alg:mc_for}.
The parametric price function that we used in line \ref{a_alg:find_cheapest_couple} is presented in Algorithm \ref{a_alg:price}.
\begin{algorithm} [htb]
\caption{Parametric Price Function}\label{a_alg:price}
\begin{algorithmic}[1]
\Procedure{Price}{$i, o$}
\begin{flushleft}
	$s$: anomaly scores based on spatial and shape anomaly measures
\end{flushleft}
  \State $d = \left(\exp{\left(\sqrt{\mbox{\sc ImpurityDist}(i,o)}\right)}\right)^{c_1}$ \label{price:d}
  \State $s = \left(1 - \left(s[i] * c_2^1 \right)^{c_3^1} * \left(s[o] * c_2^2 \right)^{c_3^2} \right)^{c_4}$ \label{price:s}
  \State $price = d * s$ \label{price:d*s}
  \If{$\exists c' \in clusters$ s.t. $o \in c'.\mathcal{C}$} \label{price:if_core}
    \State $dis = \left(1 - \left(s[i] * c_2^3 \right)^{c_5^1} * \left(s[o] * c_2^4 \right)^{c_5^2} \right)^{c_6}$ \label{price:discount}
    \State $pen = \left(2 - \left|s[i] - s[o] \right| \right)^{c_7}$ \label{price:penalty}
    \State $price = price * dis * pen$ \label{price:discount*penalty}
  \EndIf
  \State \textbf{return} $price$
 \EndProcedure
\end{algorithmic}
\end{algorithm}

\begin{figure} [htb]
\centering
\begin{subfigure}[b]{0.4\linewidth}
    \begin{tikzpicture}[scale=0.4]
    \begin{axis}[
        xtick={0, 1},
        ztick={0, 1},
        x tick label style={inner sep=0, font=\large,},
        z tick label style={inner sep=0, font=\large,},
        xtick pos=left,
        ztick pos=left,
        ymajorticks=false
    ]
    \addplot3[
        mesh,
        samples=20,
        domain=0:1,
        colormap/jet,
    ]
    {(1-(x*0.95)^0.5*(y*0.95)^0.5)^1.6};
    
    \end{axis}
    \end{tikzpicture}
    \caption{$s$ in Alg. \ref{a_alg:price} }\label{fig:price_s}
\end{subfigure}
\hspace{1.5pt}
\begin{subfigure}[b]{0.4\linewidth}
    \begin{tikzpicture}[scale=0.4]
    \begin{axis}[
        xtick={0, 1},
        ztick={0, 1},
        x tick label style={inner sep=0, font=\large,},
        z tick label style={inner sep=0, font=\large,},
        xtick pos=left,
        ztick pos=left,
        ymajorticks=false
    ]
    \addplot3[
        mesh,
        samples=20,
        domain=0:1,
        colormap/jet,
    ]
    {(1-(x*0.95)^0.05*(y*0.95)^0.05)^2.5};
    \end{axis}
    \end{tikzpicture}
    \caption{$dis$ in Alg. \ref{a_alg:price}}\label{fig:price_discount}
\end{subfigure}
\caption{Lines \ref{price:s}, \ref{price:discount} in Alg. \ref{a_alg:price}. Line \ref{price:s} imposes a price discount for all impurities, and line \ref{price:discount} imposes a much higher price discount for core impurities, both based on their anomaly scores.}
\label{fig:anomaly_scores_function}
\end{figure}
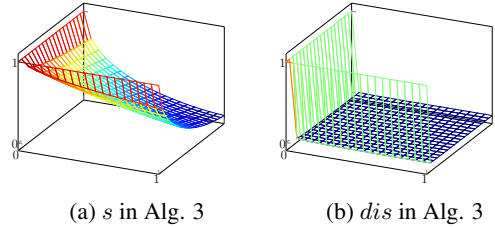

The parameters that we used are: $c_1 = 1.7,$ for $i \in [1-4]$ $c_2^i=0.95,$ $c_3^1=c_3^2=0.5,$ $c_4=1.6,$ $c_5^1=c_5^2=0.05,$ $c_6=2.5,$ $c_7=8$. The price between two impurities $i,o$ is determined by a function of the distance between them in line \ref{price:d} and by a function of how anomalous are they in line \ref{price:s}. $d$ grows with the distance, i.e. the price gets higher as the impurities are more distant from each other. However, $s$ scales negatively with the anomaly scores of $i$ and $o$. The behavior of the function can be seen at Fig. \ref{fig:price_s}. 
Since we want clusters to merge (line \ref{a_alg:other_is_core} in Algorithm \ref{mc}) and span a larger anomalous area, we check in line \ref{price:if_core} if $o$ is a core impurity of some cluster, and if it is there is a price reduction in line \ref{price:discount}. This price reduction is similar to line \ref{price:s}, yet the value $dis$ falls much more drastically than $s$. This is because we encourage clusters merging, therefore give a relatively low price to core impurities despite the distance to them. The behavior of the function in line \ref{price:discount} is presented in Fig. \ref{fig:price_discount}. Line \ref{price:penalty} penalizes cluster merging of similar sizes, in order to encourage big and anomalous clusters to absorb smaller and less anomalous clusters in cheaper price.

\begin{figure*}[ht]
\centering
\begin{subfigure}[b]{.4\linewidth}
\centering
\includegraphics[width=1.0\linewidth, frame]{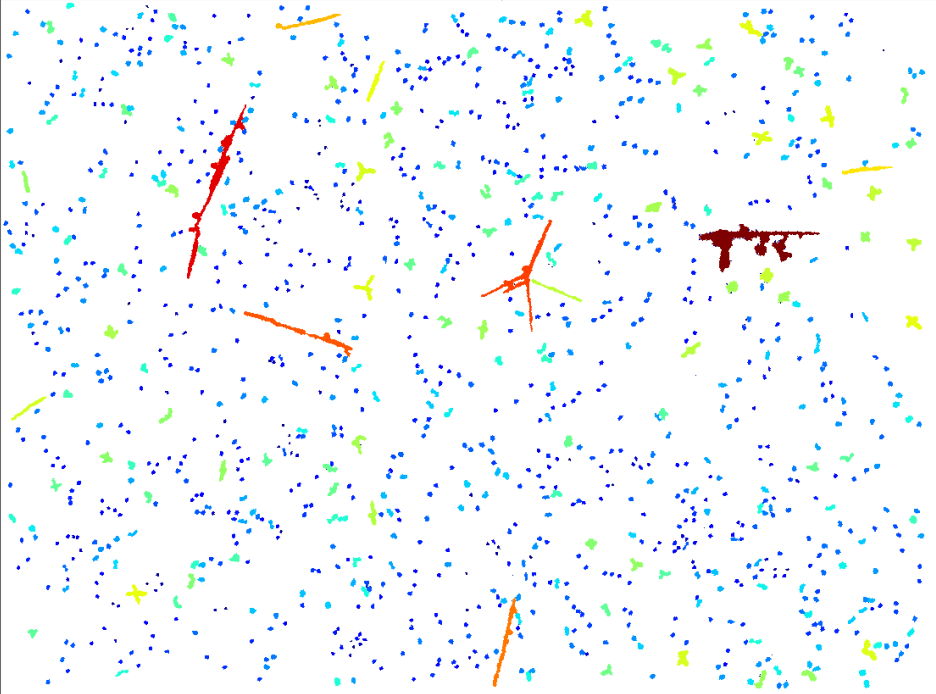}
\caption{The output of the spatial anomaly detection algorithm with $k=50$. The anomaly scores are normalized to $[0,1]$, while the most anomalous impurities are with scores close to 1 and are colored in red, and the most non-anomalous impurities are with scores closed to 0 and are colored in blue.\newline} \label{fig:spatial}
\end{subfigure}
\hspace{0.25cm}
\begin{subfigure}[b]{.4\linewidth}
\centering
\includegraphics[width=1.0\linewidth, frame]{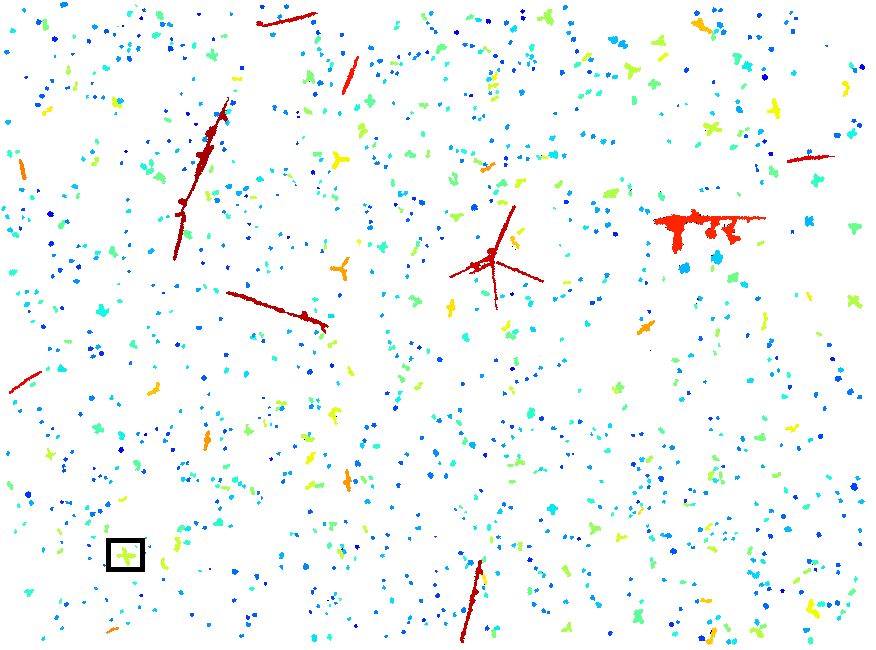}
\caption{Measuring shape anomaly score of each impurity  according to Equation  \ref{eq_circle_diff}.
In the black rectangle there is an example of anomalous 'X' shaped impurity in the middle of the shape anomaly scale. This impurity should get a higher shape anomaly score.
\newline\newline} \label{fig:circle_diff}
\end{subfigure}
\hspace{0.08cm}
\begin{subfigure}[b]{.05\linewidth}
\centering
\includegraphics[height=150pt]{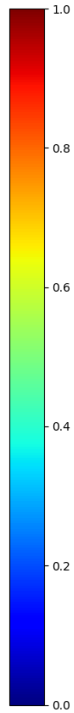}
\caption*{
\newline \newline \newline \newline \newline\newline}
\end{subfigure}
\begin{subfigure}[b]{.4\linewidth}
\centering
\includegraphics[width=1.0\linewidth, frame]{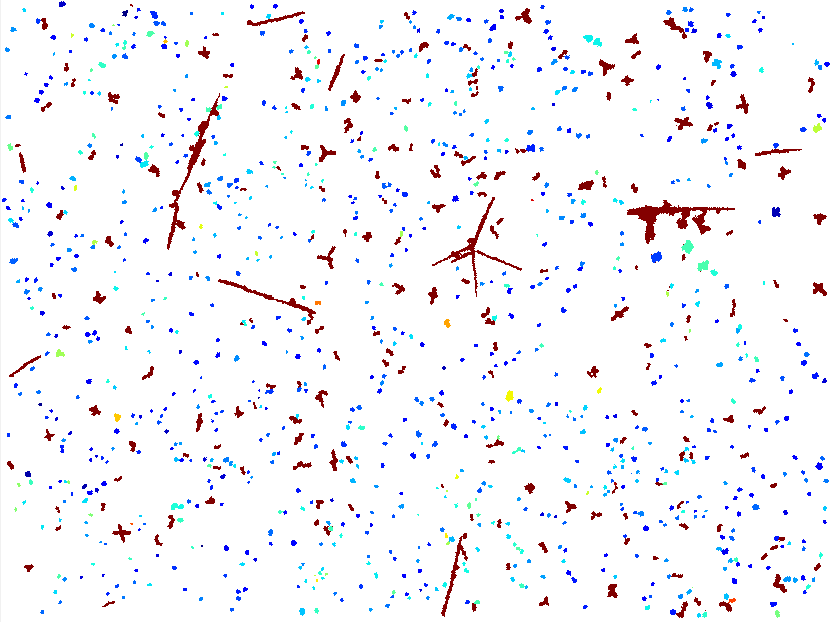}
\caption{The output of the model that uses blank images as labels for the anomalous impurities in the training phase of the AE.\newline} \label{fig:shape_blank}
\end{subfigure}
\hspace{0.25cm}
\begin{subfigure}[b]{.4\linewidth}
\centering
\includegraphics[width=1.0\linewidth, frame]{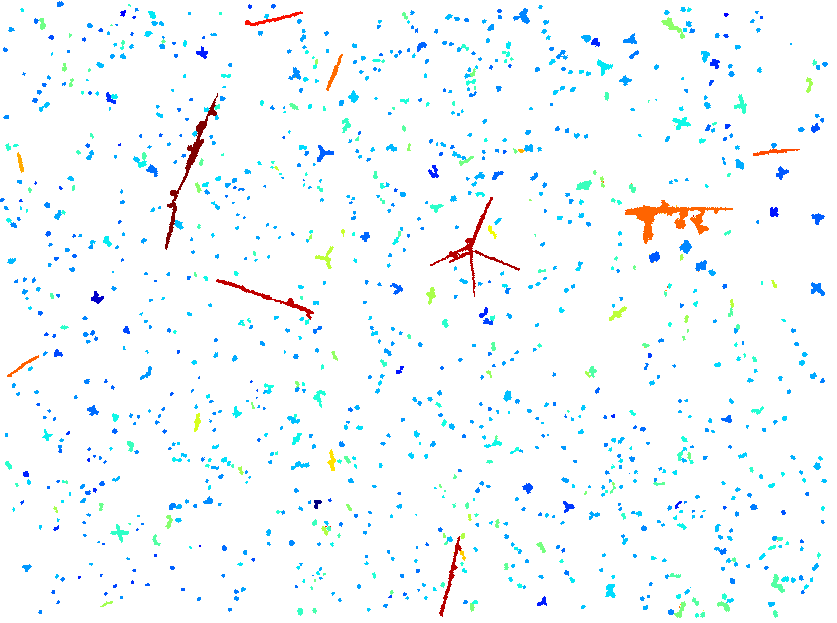}
\caption{The output of the model that uses only normal impurities in the training phase of the AE.\newline \newline} \label{fig:shape_same}
\end{subfigure}
\hspace{0.08cm}
\begin{subfigure}[b]{.05\linewidth}
\centering
\includegraphics[height=150pt]{colormap.PNG}
\caption*{
\newline \newline \newline}
\end{subfigure}
\begin{subfigure}[b]{.4\linewidth}
\centering
\includegraphics[width=1.0\linewidth, frame]{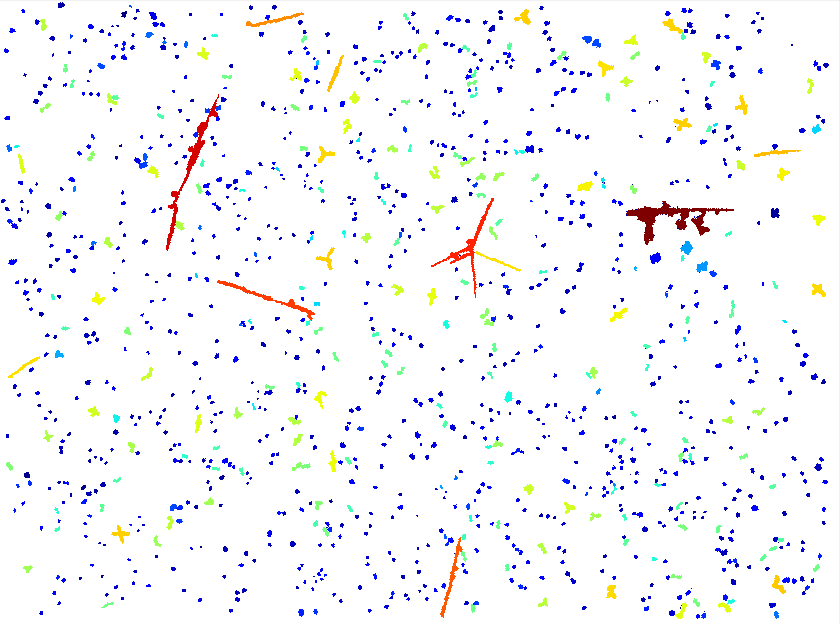}
\caption{The combined measure of shape and spatial anomaly. \newline} \label{fig:shape_spatial}
\end{subfigure}
\hspace{0.25cm}
\begin{subfigure}[b]{.4\linewidth}
\centering
\includegraphics[width=1.0\linewidth, frame]{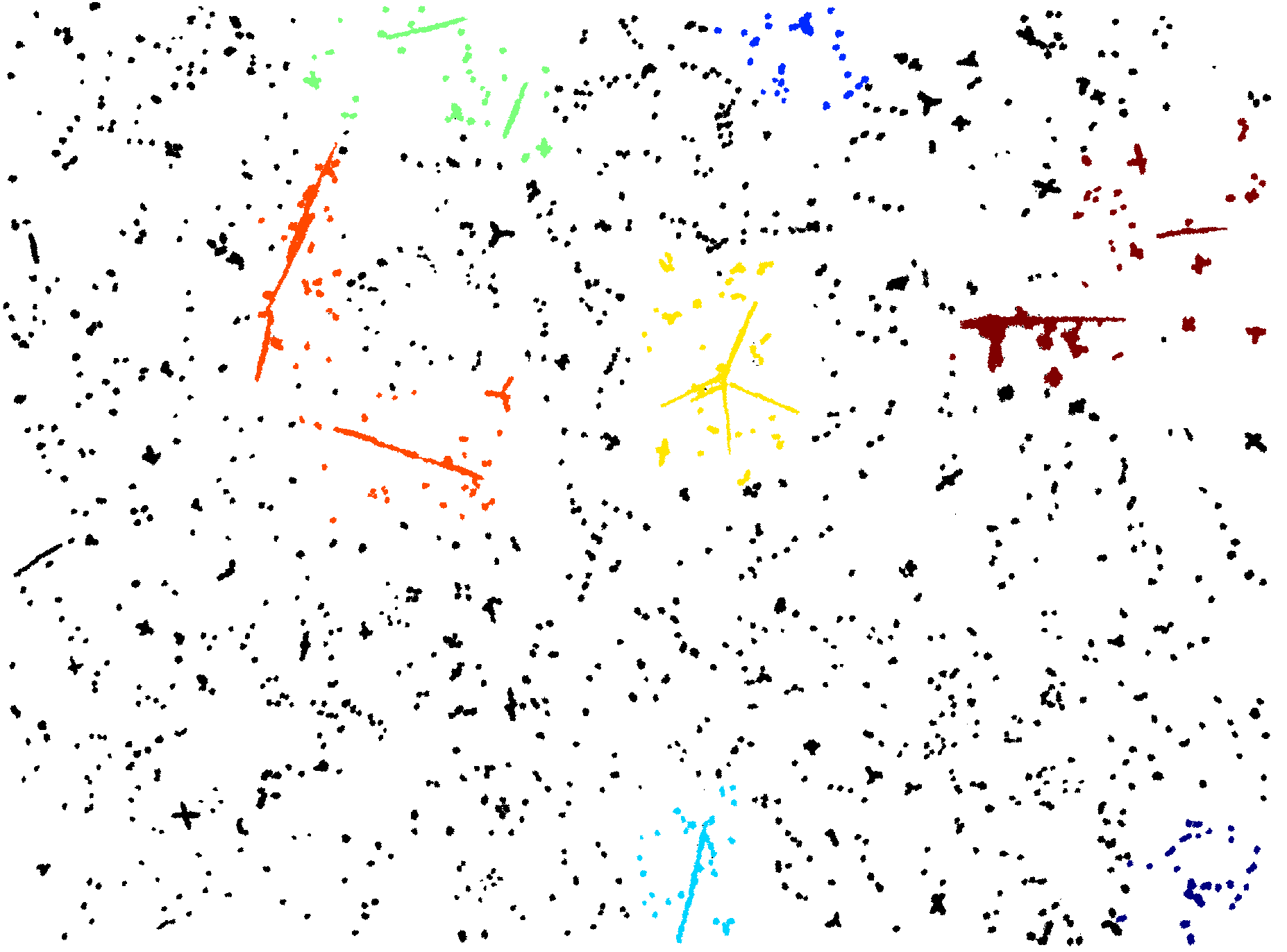}
\caption{The output of the Market Clustering algorithm as the area anomaly measure, with $k=10$, ordered by Equation \ref{eq_area_measure}. The most anomalous cluster is the red one.} \label{fig:area}
\end{subfigure}
\hspace{0.08cm}
\begin{subfigure}[b]{.05\linewidth}
\centering
\includegraphics[height=150pt]{colormap.PNG}
\caption*{
\newline \newline}
\end{subfigure}
\caption{Plots of the algorithms on an example scan form the data-set.} \label{fig:all_plots}
\end{figure*}
After marking the anomalous areas, we now quantify the degree of anomaly of each area. We suggest the following area anomaly measure for cluster $c$, and we next prove that it indeed indicates how $c$ is anomalous.
\begin{theorem}
$am(c)$ is monotonically increasing with the degree of anomaly of $c$, based on \mbox{\sc MarketClustering} algorithm and on $c$'s Spatial and Shape anomaly score, where $am(c)$ is defined as:
\end{theorem}
\vspace{-0.5cm}
\begin{equation}
\sum_{i\in c.\mathcal{I}} \left(\mbox{\sc Score}(i)*\mbox{\sc Area}^{2}(i)\right) * \mbox{\sc Diameter}(c) * |c.\mathcal{I}| \label{eq_area_measure}
\end{equation} 
\begin{proof}
Appending lots of non-anomalous (and not core) impurities is an expensive procedure, compared to appending lots of anomalous impurities, because of the discount in line \ref{price:s} in Algorithm \ref{a_alg:price} for anomalous impurities. Thus, clusters with a large number of impurities apparently have included cheaper, more anomalous, impurities. Moreover, clusters that have appended core impurities, which are the most anomalous impurities, clearly should be considered as more anomalous. Indeed, cluster merging imposes higher wallet balance for future impurities addition, in addition to a concatenation of impurities in each cluster. 
Thus, as the amount of impurities in the cluster, $|c.\mathcal{I}|$, grows, the degree of anomaly of $c$ grows correspondingly.
Similarly, appending some far impurity $o$ (line \ref{price:d} in Algorithm \ref{a_alg:price}) is naturally an expensive operation, as long as $o$ is not anomalous (line \ref{price:s}). Thus, if a cluster overcame the expenses of appending distant impurities, it is probably because it has appended anomalous impurities. Therefore, as $\mbox{\sc Diameter}(c)$ grows, the degree of anomaly of $c$ grows as well.
Finally, big and anomalous impurities highly imply that the cluster has a high spatial anomaly score. Therefore the component $\sum_{i\in c.\mathcal{I}} \left(\mbox{\sc Score}(i)*\mbox{\sc Area}^{2}(i)\right)$ grows with the degree of anomaly of $c$.

Since all components are monotonically increasing with the degree of anomaly of $c$, and because multiplication preserves monotonicity, $am(c)$ is monotonically increasing with the degree of anomaly of $c$.
\end{proof}
We also note that, similarly to all other anomaly measures presented in this work, the usage of multiplication enhances the anomaly scores of clusters with high scores in each of the components, compared to clusters with lower scores on some of the components.
The output of \mbox{\sc MarketClustering} on top of the spatial and shape anomaly measures, and after ordering the clusters based on Equation \ref{eq_area_measure}, is presented in Fig. \ref{fig:area}.

\section{Experiments}
\begin{figure}[htb]
\centering
\begin{subfigure}[b]{.45\linewidth}
\centering
\includegraphics[height=2.7cm, frame]{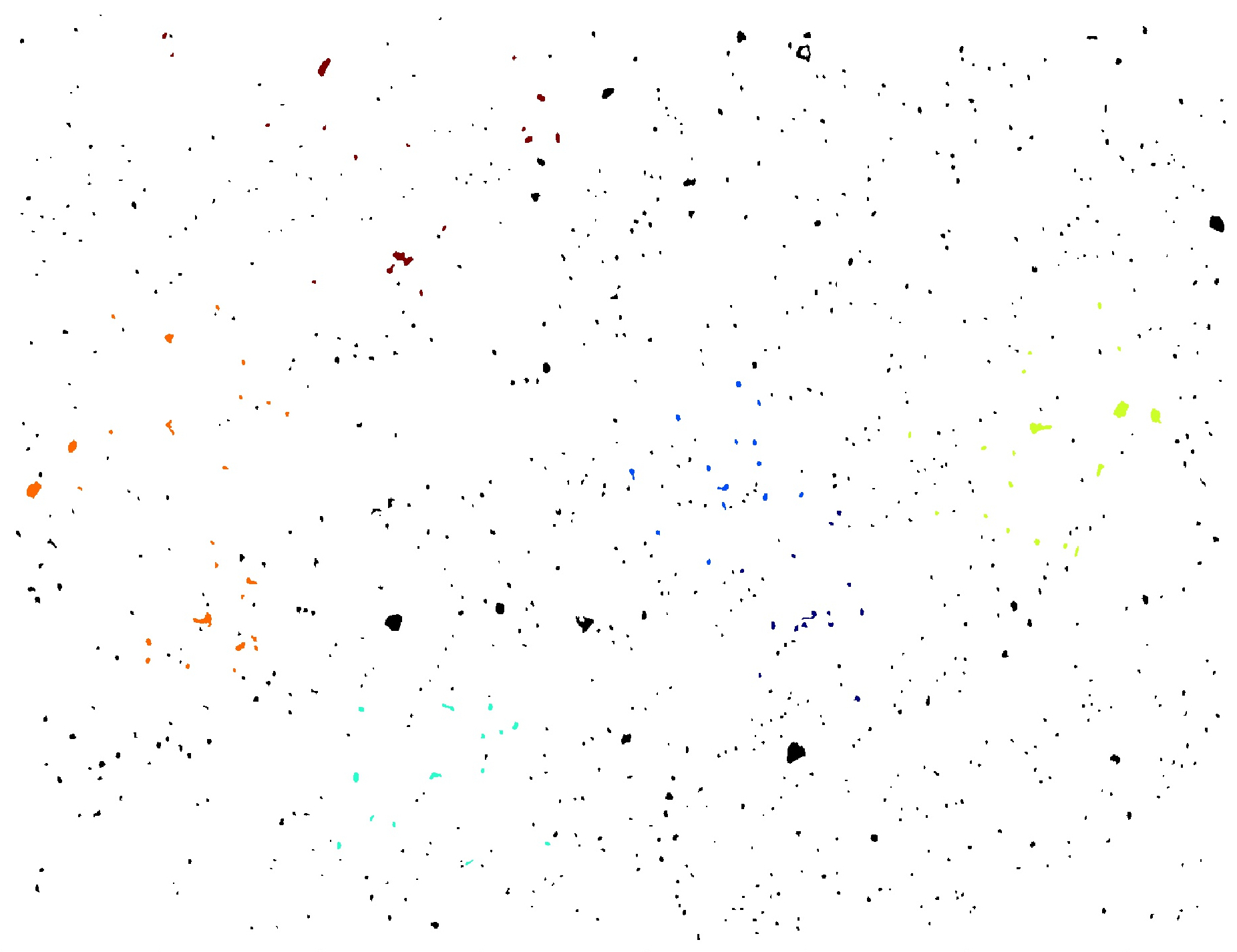}
\caption{The output of the model on $img_1$.} \label{fig:out_img_1}
\end{subfigure}
\begin{subfigure}[b]{.45\linewidth}
\centering
\includegraphics[height=2.7cm, frame]{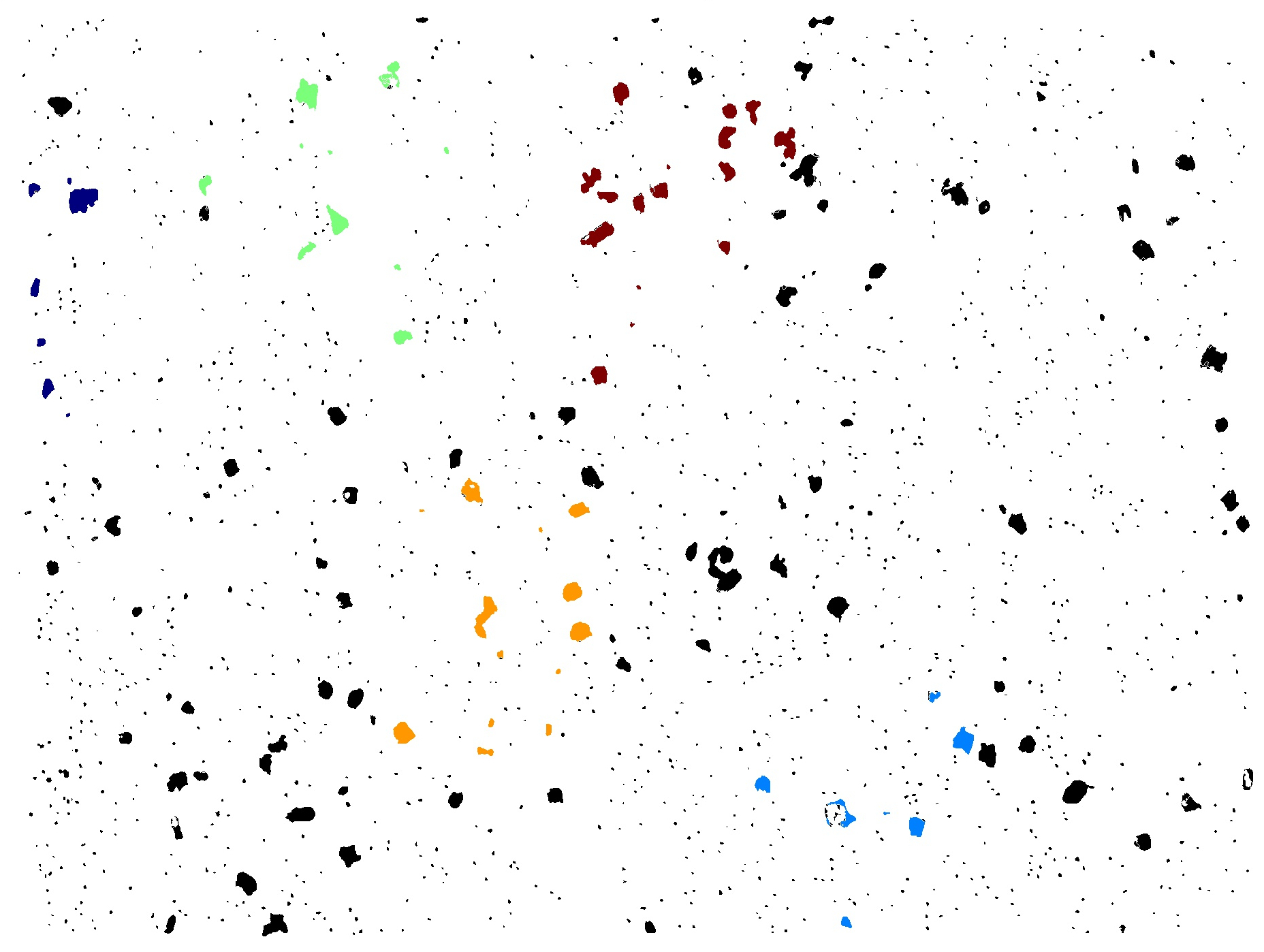}
\caption{The output of the model on $img_3$.} \label{fig:out_anom}
\end{subfigure}
\begin{subfigure}[b]{.05\linewidth}
\centering
\includegraphics[height=2.75cm]{colormap.PNG}
\caption*{\newline}
\end{subfigure}
\caption{Plots of the whole model on an test scans.} \label{fig:all_tests}
\end{figure}

In order to correlate the model results with physical measure, we used the following procedure: We prepared three fresh metallographic samples, and used the model to locate and quantify the anomaly scores of the most anomalous areas of impurities inside them. The outputs of the model are then used in order to determine whether and where there were physical defects in the materials, specifically in the areas of interest. The results are shown in Figures \ref{fig:out_img_1} and \ref{fig:out_anom}. We omitted from space considerations the result for $img_2$, since it highly resembles $img_1$. The anomalous areas of the test scans were ordered together with all other scans in the data-set, and they were placed in places 1588, 1642 and 916 respectively, out of totally 1653 clusters. Clusters 1588 and 1642 resulted under the first decile, while 916 was placed in between the fourth and fifth deciles. All the results are shown in Fig. \ref{fig:all_ranks}.
After getting the outputs from the model, two examinations were made: 1. Microhardness Vickers (MHV) \cite{buckle1959progress} test in the vicinity of the most anomalous inclusions (the inclusions size and the microhardness trace are both on the same scale of few to tens micrometers) and in normal areas; 2. EDS (energy dispersive spectroscopy) analysis in a SEM \cite{goldstein2017scanning} (scanning electron microscope) to evaluate the inclusions composition.
For the three samples that were examined by MHV and EDS, within the sensitivity and the accuracy of these methods, there was no difference between normal and anomalous inclusions.
Although it is clear that the most anomalous area in \ref{fig:out_anom} looks much more anomalous than the others, and since we observed that there is no difference between normal and anomalous areas, we conclude that our computer model is capable to quantify successfully how each area of inclusions is anomalous compared to all other areas, but it is the task of the experienced user of the system to determine the threshold from which the area is considered anomalous enough to be defective.

\pgfplotsset{width=7cm,compat=1.8}
\begin{figure}[htb]
\centering
\resizebox{0.45\textwidth}{!}{
\begin{tikzpicture}
    \begin{axis}[
        ybar=0.85pt,
        height=60mm, 
        width=60mm,
        xlabel= clusters,
        ylabel= ranks,
        bar width=4pt,
        x = 0.5cm,
        legend pos=outer north east,
        legend entries={0\%-10\%,
              20\%-30\% ,
              30\%-40\% ,
              40\%-50\%
              },
        enlarge x limits={abs=2cm},
        ymin=0,
        ytick={916,1117, 1193, 1530, 1650},
        xtick={-2.2,1.1,4.5},
        xticklabels={{$img_1$},{$img_2$},{$img_3$}},
        y tick label style={inner sep=0, font=\scriptsize,},
        x tick label style={align=center,font=\footnotesize,},
        xtick pos=left,
        ytick pos=left,
    ]
    \addlegendimage{fill=blue}
    \addlegendimage{fill=green}
    \addlegendimage{fill=orange}
    \addlegendimage{fill=red}
    \addplot [fill=blue] coordinates {(0,1650)};
    \addplot [fill=blue] coordinates {(0,1649)};
    \addplot [fill=blue] coordinates {(0,1646)};
    \addplot [fill=blue] coordinates {(0,1621)};
    \addplot [fill=blue] coordinates {(0,1602)};
    \addplot [fill=blue] coordinates {(0,1588)};
    \addplot [fill=blue] coordinates {(1,1656)};
    \addplot [fill=blue] coordinates {(1,1655)};
    \addplot [fill=blue] coordinates {(1,1654)};
    \addplot [fill=blue] coordinates {(1,1653)};
    \addplot [fill=blue] coordinates {(1,1652)};
    \addplot [fill=blue] coordinates {(1,1651)};
    \addplot [fill=blue] coordinates {(1,1647)};
    \addplot [fill=blue] coordinates {(1,1642)};
    \addplot [fill=blue] coordinates {(2,1552)};
    \addplot [fill=blue] coordinates {(2,1531)};
    \addplot [fill=green] coordinates {(2,1193)};
    \addplot [fill=orange] coordinates {(2,1117)};
    \addplot [fill=red] coordinates {(2,916)};
    \end{axis}
\end{tikzpicture}
}
\caption{Ranks of all clusters, sperated to groups, such that each group is a single test scan.} \label{fig:all_ranks}
\end{figure}

\section{Conclusions}
In this study we presented a comprehensive approach for anomaly detection in geometric objects (impurities), using three measures: Spatial, Shape and Area anomaly. The first two are used to measure the degree of anomaly of each object, and the third for marking and quantifying the degree of anomaly of areas of objects.
Whenever the anomaly score of an area gets lower than some threshold determined by an expert, the sample can be approved as sound with no further examination. However, when it gets higher score, it should be examined by the methods suggested here (MHV and EDS), or by other relevant experimental method. If the results show that there is no difference between normal and anomalous inclusions (as looks can deceive), the threshold score for sound samples can be updated, but if there is a difference, the sample must be suspended for further and deeper examination.

\bibliographystyle{ieee_fullname}
\bibliography{egbib}

\end{document}